\documentclass[aps,11pt,showpacs,onecolumn,nofootinbib]{revtex4}

\usepackage[utf8]{inputenc}
\usepackage{array}
\usepackage[english]{babel}
\usepackage{amsmath}
\usepackage{amsfonts}
\usepackage{amssymb}

\usepackage{hyperref}
\usepackage{subeqnarray}
\usepackage{tensor}

\usepackage{graphicx}
\usepackage{multirow}

\usepackage{color}

\DeclareMathOperator{\I}{i}
\DeclareMathOperator{\E}{e}

\begin{document}

\title{A nearly cylindrically symmetric source in the Brans-Dicke gravity as the generator of the rotational curves of the galaxies}

\author{S. Mittmann dos Santos}
\email{sergio.santos@poa.ifrs.edu.br}
\affiliation{Universidade Estadual Paulista Júlio de Mesquita Filho – UNESP \\
Campus de Guaratinguetá, 12516-410, Guaratinguetá, SP, Brazil and \\
Instituto Federal de Educação, Ciência e Tecnologia do Rio Grande do Sul – IFRS \\
Campus Porto Alegre, 90030-041, Porto Alegre, RS, Brazil}

\author{J. M. Hoff da Silva}
\email{hoff@feg.unesp.br}
\affiliation{Universidade Estadual Paulista Júlio de Mesquita Filho – UNESP \\
Campus de Guaratinguetá, 12516-410, Guaratinguetá, SP, Brazil}

\author{M. E. X. Guimar\~aes}
\email{emilia@if.uff.br}
\affiliation{Instituto de F\'{\i}sica, Universidade Federal Fluminense - UFF \\
Av. Gal. Milton Tavares de Souza, s/n, 24210-346, Niter\'oi, RJ, Brazil}

\author{J. L. Neto}
\email{jlneto@if.ufrj.br}
\affiliation{Instituto de F\'{\i}sica, Universidade Federal do Rio de Janeiro - UFRJ \\
Av. Athos da Silveira Ramos, 149, 21941-909, Cidade Universitária, Rio de Janeiro, RJ, Brazil}

\begin{abstract}
Observation shows that the velocities of stars grow by approximately 2 to 3 orders of  magnitude when
the distances from  the centers of
the galaxies are in the range of $0.5$ kpc to $82.3$ kpc, before they begin to tend to a constant value. Up to know, the reason for this behavior is still a matter for debate. In this work, we propose a model which  adequately describes this unusual behavior  using a (nearly) cylindrical symmetrical solution in the framework of a scalar-tensor-like (the Brans-Dicke model) theory of gravity.
\end{abstract}


\maketitle

\section{Introduction}

Brans-Dicke (BD) gravitation \cite{Brans1961} describes gravitational phenomena with the aid of a scalar field $\tilde{\phi}$ in addiction to the usual tensorial field in General Relativity (GR). The gravitational action of the BD theory, which gives rise to the dynamic equations, is given by (in the Jordan-Fierz physical frame)\footnote{We have reserved the terms with tilde for the evaluated quantities in the Jordan-Fierz frame.}
\begin{equation}
S=\frac{1}{16 \pi} \int d^4 x \sqrt{-\tilde{g}} \left ( \tilde{\phi} \tilde{R} -\frac{\omega}{\tilde{\phi}} \, \partial^\mu \tilde{\phi} \partial_\mu \tilde{\phi} \right )+ S_\textit{matter} \; .
\label{acao_Brans-Dicke}
\end{equation}
The dimensionless parameter $\omega$, called the BD coupling parameter, determines the deviation of the results obtained in this theory from those in GR.
The lower the value of $\omega$, the more different are the correspondent results.
For instance, for the Solar System, where GR does extremely well, $|\omega|>40000$ \cite{limite_experimental}.
Thus, one could think of discarding a scalar-tensor-like theory.
Nevertheless, during the investigation of the cosmological evolution of the Universe with a scalar-tensorial theory it was concluded that GR is  an attractor of the scalar-tensor theory, because the scalar field dynamics would have been gradually suppressed \cite{Damour1993, Damour1993A, Polyakov1994, Contaldi1999}.
Therefore, a scalar-tensor gravitation theory seems to be most appropriate for describing sources that originated shortly after the Big Bang, when scalar field strength and its variation were still important.

It is now known that only less than 5$\%$  of what makes up the Universe is well understood, consisting of baryonic matter and radiation. Dark matter and dark energy,  which dominate our Universe,  have not yet their origin and evolution satisfactorily explained. Dark matter, supposedly  present in the halo region of galaxies, seems to be the mechanism that causes the unusual behavior of the tangential velocity of the stars which is superior to that predicted by interaction with visible matter \cite{Zwicky1933, Zwicky1937, Rubin1980, Persic1995, Persic1997}. Moreover, in addition to the velocity larger than expected, as the distance of the stars from the center of the galaxy grows, their velocities tend to a constant value \cite{Rubin1980}.
An alternative explanation to that behavior is to consider a scalar-tensor theory which might solve the rotational curves paradigm \cite{Guzman2002, Lee2004, Leineker2006}. The results of the abovementioned references do reproduce the large distance behavior of the velocity with the expected order of magnitude. However, they give no information on how the intensity of that velocity evolves.

The stars of the galaxies studied in Ref. \cite{Rubin1980} are distributed along an approximately flat region.
Therefore, it is possible to suppose that a symmetrically cylindrical spacetime could describe the rotation curves.
Symmetrically cylindrical sources are widely studied in several areas.
Among these sources, some prominence can be given to the spinning cosmic strings, which would have originated from a spontaneous symmetry breaking, occurring from a phase transition of the primordial Universe.
They have an angular velocity about the longitudinal axis of symmetry and, being straight strings, they preserve the Lorentz invariance along its  symmetry axis.
Many studies in the GR theory indicate that there are closed timelike curves (CTC's) at least in part of the spacetime around these strings \cite{Jensen1992}.
As they would have formed in the primordial Universe, a recent work has demonstrated what occurs when, instead of GR,  the description of these defects is analysed in the framework a BD gravity \cite{artigo_corda_reta}.
As in GR, the results indicated that CTC's are also present in the spacetime around these strings.
There is a  great deal of discussion as to whether CTC's exist or not in the Universe,
since their confirmation would violate the causality principle.
It is argued that some unknown form of matter could allow the existence of CTC's \cite{Deser1992}, which
can not be ruled out, since we do not have a full understanding of what most of the Universe is made up.
The works \cite{artigo_corda_reta, Mazur1986} analyzed the energy quantization of a particle with a non inertial motion around a straight spinning cosmic string and concluded that it would be unstable, because it would have an angular velocity much higher than would be physically acceptable.
Thus, if there was the formation of this type of string in the primordial Universe, it should not have survived.

The main goal of this paper is to present the full solution for a nearly cylindrically symmetric spacetime in the BD gravity and use it, as an alternative to dark matter, to describe the rotational curves of the galaxies in Ref. \cite{Rubin1980}.
We show the solutions for the equations of motion for the  action (\ref{acao_Brans-Dicke}) when Lorentz invariance along the $z$ longitudinal axis of symmetry is slightly violated, given that a straight source may be unstable, as happen with the spinning cosmic strings.
The relevant invariance when dealing with the gravitational phenomena is to be taken with respect to general coordinate transformations. Since we are interested in the vacuum solution, less importance is devoted to the source of this spacetime.
Nevertheless, one may bear in mind a (non-straight) spinning cosmic string with a very large radius.
If one thinks of the source as a Lorentz violating object, then it is important to remember that many works have pointed out that the search for a theory of grand unification leads to a spontaneous breaking of Lorentz symmetry at the time of the primordial Universe \cite{Kostelecky1989, Kostelecky1991},  when we consider that the formation of the source has occurred and, therefore, also the BD gravity was dominant over GR.
Moreover, in order not to lose the (nearly) cylindrical symmetry adopted, the ratio between the $dt^2$ and $dz^2$ coefficients of the line element is approximately unity.

We rescale the equations of motion obtained from the action in the Jordan-Fierz frame (\ref{acao_Brans-Dicke}) to the Einstein (or conformal) frame in which the scalar and the tensor degrees of freedom do not mix. This makes the interpretation and handling of the solutions easier. It is important to strength that the qualitative results obtained in one frame are equivalent to those in the other.
In any case, throughout the paper the interpretation of the results will be performed for the two frames, since both cases may be important in the description of a given gravitational system \cite{Faraoni1999, Quiros2013}. With the solutions of the dynamic equations at hand, we verified that the spacetime is indeed regular as evidenced by the analysis of the Kretschmann scalars. Depending on the values adopted for the integration constants, CTC's may appear  or not in certain regions. The  resulting spacetime adequately reproduces the motion of the stars of the galaxies studied in Ref. \cite{Rubin1980}, either in the regions where  CTC's exist or where the metric is Euclidean. This is the main result of our paper. We also provide additional argumentation in favor of using  BD gravity to address this problem.

This paper is organized as follows. In Section \ref{Secao2},  we first present the dynamic equations for the BD action in the Jordan-Fierz and Einstein frames.
From the equations in the Einstein frame, the exact solution for a nearly cylindrically symmetric spacetime is obtained. In Section \ref{Secao3}, the physical properties of the spacetime obtained in the previous section are described. In Section \ref{Secao4}, we show that the resulting spacetime adequately describes the rotational curves of Sc type galaxies. In Section \ref{Secao5}, it is discussed that the BD gravitation is essential for the aforementioned application.
In Section \ref{Secao6}, we summarize our findings and discuss the relevant conclusions. We leave for the Appendix complete tables encompassing relevant data and calculation results of the presented analysis.

\section{Solutions for a nearly cylindrically symmetric spacetime in Brans-Dicke gravity}
\label{Secao2}

The equations of motion from the action (\ref{acao_Brans-Dicke}) in the Jordan-Fierz frame are
\begin{eqnarray}
\tilde{G}_{\mu\nu}&=&\frac{\omega}{\tilde{\phi}^{\,2}} \left ( \partial_\mu \tilde{\phi} \partial_\nu \tilde{\phi} - \frac 1 2 \tilde{g}_{\mu\nu} \tilde{g}^{\rho \sigma} \partial_{\rho} \tilde{\phi} \partial_\sigma \tilde{\phi} \right ) +\frac 1 {\tilde{\phi}} \left ( \partial_\mu \partial_\nu \tilde{\phi} - \tilde{g}_{\mu\nu} \tilde{\Box} \tilde{\phi} \right )+ \frac{8\pi}{\tilde{\phi}}T_{\mu\nu} \; , \nonumber \\
\tilde{\Box} \tilde{\phi} &=& \frac 1 {\sqrt{-\tilde{g}}} \; \partial_\mu \left ( \sqrt{-\tilde{g}} \; \partial^\mu \tilde{\phi} \right )= \frac {8\pi}{2\omega +3} T \; .
\label{eqs_movimento_BD_Jordan}
\end{eqnarray}
There is a more convenient way to handle  the BD equations which is through the Einstein frame. In this frame, $\tilde{g}_{\mu\nu}$ and $\tilde{\phi}$ from the Eqs. (\ref{eqs_movimento_BD_Jordan}) are rescaling by the new dynamical variables $g_{\mu\nu}$ and $\phi$, as in \cite{Boisseau1998}:
\begin{eqnarray}
\tilde{g}_{\mu\nu}& = & \exp \left( 2\kappa \phi\right) g_{\mu\nu} \; ,\label{variaveis_Jordan-Einstein_1} \\
\tilde{\phi}& = & \frac 1 G \exp \left( -2 \kappa \phi \right) \; ,
\label{variaveis_Jordan-Einstein_2}
\end{eqnarray}
where
\begin{equation}
\kappa^2=\frac 1 {2 \omega + 3}
\label{kappa}
\end{equation} and $G$ is the effective gravitational Newtonian constant. Thus, in the Einstein frame, Eqs. (\ref{eqs_movimento_BD_Jordan}) can be written as
\begin{eqnarray}
G_{\mu\nu}&=& 2 \partial_\mu \phi \partial_\nu \phi - g_{\mu\nu} g^{\rho \sigma} \partial_{\rho} \phi \partial_\sigma \phi + \frac{8\pi}{\phi}T_{\mu\nu} \; , \label{eqs_movimento_BD_Einstein_1}
 \\
\Box \phi &=&
%\frac 1 {\sqrt{-g}} \partial_\mu \left ( \sqrt{-g} \partial^\mu \phi \right ) =
\frac {8\pi}{2\omega +3} T \; .
\label{eqs_movimento_BD_Einstein_2}
\end{eqnarray}
For the system we are about to consider, since the qualitative behavior is basically the same in Jordan-Fierz and Einstein frames, we shall present and interpret the results in the manageable Einstein frame and only occasionally make reference to the Jordan-Fierz frame.

The most general cylindrically symmetric line element to be addressed in this work is \cite{Jensen1992}
%, artigo_corda_reta}
\begin{equation}
ds^2 = -\left( \E^\alpha dt+M \; d\varphi \right)^2+r^2 \E^{-2\alpha} d\varphi^2 + \E^{2 \left( \beta-\alpha \right) } \left( dr^2+dz^2 \right) \; ,
\label{metrica}
\end{equation}
where the angular coordinate varies, in principle, in the range $0\leq\varphi < 2\pi$.
By means of the cylindrical symmetry we have $\alpha$, $\beta$ and $M$ as functions of the, strictly positive, radial distance $r$ only.
In order not to completely give up  cylindrical symmetry we adopt $\frac{\beta}{\alpha} \approx 2$.
It is in this sense that we mean nearly cylindrical symmetry. Notice that an eventual source for this
spacetime would not be straight and slightly breaking
Lorentz invariance. The field equations were obtained in detail previously \cite{artigo_corda_reta}.
Therefore we will suppress the calculations. They read, for the vacuum case,
\begin{eqnarray}
\mathcal{G}^t_{\;\;t}&:&-\;3\Omega^2-\left( 2\alpha''-\beta''+\frac{2\alpha'}{r}-\alpha'^2 \right) \E^{2 \left( \alpha-\beta \right)}
=-\;\E^{2 \left( \alpha-\beta \right)} \phi'^2
\label{sistema_Gtt} \; , \\
\mathcal{G}^r_{\;\;r}&:&\Omega^2+\left( \frac {\beta'}{r}-\alpha'^2 \right) \E^{2 \left( \alpha-\beta \right)}
=  \E^{2 \left( \alpha-\beta \right)} \phi'^2
\label{sistema_Grr} \; , \\
\mathcal{G}^z_{\;\;z}&:&-\;\Omega^2+\left( \alpha'^2-\frac{\beta'}{r} \right) \E^{2 \left( \alpha-\beta \right)}
= -\;\E^{2 \left( \alpha-\beta \right)} \phi'^2
\label{sistema_Gzz} \; , \\
\mathcal{G}^\varphi_{\;\;\varphi}&:&\Omega^2+\left( \beta''+\alpha'^2 \right) \E^{2 \left( \alpha-\beta \right)}
= -\;\E^{2 \left( \alpha-\beta \right)} \phi'^2
\label{sistema_Gvarphivarphi} \; , \\
\mathcal{G}^t_{\;\;\varphi}&:&\left [ \Omega'+2\alpha' \Omega+\Omega \left( \beta'-\alpha' \right) \right ] \E^{\alpha-\beta}
= 0
\label{sistema_Gtvarphi} \; , \\
\Box \phi &:& \frac {\phi'} 2 \frac {d}{dr} \left\{ \ln \left[\E^{4\left( \beta-\alpha \right)} r^2 \left( \E^{2 \left( \alpha-\beta \right)} \phi' \right)^2\right] \right\} \E^{2 \left( \alpha-\beta \right)}
= 0 \; ,
\label{sistema_Boxphi}
\end{eqnarray}
where
\begin{equation}
\label{Omega}
\Omega = \frac{M\alpha'-M'}{2r}\; e^{2\alpha - \beta} \, .
\end{equation}
The prime $(')$ in the above equations refers to derivative with respect to $r$.

Eqs. (\ref{sistema_Gzz}) and (\ref{sistema_Grr}) are equal and Eq. (\ref{sistema_Boxphi}) has the following
 solution for the scalar field
\begin{equation}
\phi(r)=C_2+C_1 \ln r \; ,
\label{solucao_phi}
\end{equation}
where $C_1$ and $C_2$ are constants. Adding Eqs. (\ref{sistema_Gtt}) and (\ref{sistema_Grr}) and using our definition of $\Omega$ (\ref{Omega}) we get
%\begin{eqnarray}
%&&\frac 1 4 \left[ \frac{3\E^{2 \alpha} \left( M'-M\alpha' \right)^2}{r^2} -4\left( -\;\frac{2\alpha'}{r}+\alpha'^2-2\alpha''+\beta'' \right) \right] \nonumber \\
%&=&\frac 1 4 \left[ -\;4\alpha'^2+\frac{\E^{2 \alpha}\left( M'-M\alpha' \right)^2}{r^2}+\frac{4\beta'}{r} \right]
%\label{Gtt+Grr}
%\end{eqnarray}
%and can be rewritten as
\begin{equation}
\E^{2\alpha}\left( M'-M\alpha' \right)^2=-2r\left( 2\alpha'-\beta'+2r\alpha''-r\beta'' \right) \; .
\label{Gtt+Grr_2}
\end{equation}
Subtracting  Eqs. (\ref{sistema_Gtt}) from (\ref{sistema_Gvarphivarphi}) and substituting the term
$\E^{2 \alpha} \left( M'-M\alpha' \right)^2$  by the right-hand side of Eq. (\ref{Gtt+Grr_2}), we have
%\begin{eqnarray}
%&&-\;\frac 1 4 \left[ \frac{3\E^{2 \alpha} \left( M'-M\alpha' \right)^2}{r^2} -4\left( -\;\frac{2\alpha'}{r}+\alpha'^2-2\alpha''+\beta'' \right) \right] \nonumber \\
%&=&\frac 1 4 \left[ \frac{\E^{2 \alpha} \left( M'-M\alpha' \right)^2}{r^2} + 4 \left( \alpha'^2 + \beta'' \right) \right] \; .
%\label{Gtt-Gvarphivarphi}
%\end{eqnarray}
%Now, substituting the term $\E^{2 \alpha} \left( M'-M\alpha' \right)^2$ from Eq. (\ref{Gtt-Gvarphivarphi}) by the right-hand side of Eq. (\ref{Gtt+Grr_2}), we have
\begin{equation}
\alpha'-\beta'+r\alpha''-r\beta''=0 \; .
\label{Gtt-Gvarphivarphi_2}
\end{equation}
If $\alpha-\beta=\gamma$, Eq. (\ref{Gtt-Gvarphivarphi_2}) can be rewritten as
\begin{equation}
\gamma'+r\gamma''=0 \; ,
\label{eq_dif_gamma}
\end{equation}
%The equation (\ref{eq_dif_gamma}) has
which has the solution
\begin{equation}
\gamma=C_b+C_a \ln r \; ,
\label{solucao_eq_dif_gamma}
\end{equation}
where $C_a$ and $C_b$ are constants.
Thus,
%\begin{equation}
%\alpha-\beta=C_b+C_a \ln r \; .
%\label{alpha-beta}
%\end{equation}
%and
\begin{equation}
\beta(r)=\alpha(r)-\left( C_b+C_a \ln r \right) \; .
\label{solucao_beta}
\end{equation}
Eq. (\ref{sistema_Gvarphivarphi}) can be rewritten as
\begin{equation}
\frac 1 4 \left[ \frac{\E^{2 \alpha} \left( M'-M\alpha' \right)^2}{r^2} + 4 \left( \alpha'^2 + \beta'' \right) \right]=-\;\phi'^2 \; .
\label{sistema_Gvarphivarphi_2}
\end{equation}
Substituting the term $\E^{2 \alpha} \left( M'-M\alpha' \right)^2$ from the Eq. (\ref{sistema_Gvarphivarphi_2})
by the right-hand side of the Eq. (\ref{Gtt+Grr_2}), we have
\begin{equation}
\phi'^2=\frac {2\alpha'-2r\alpha'^2-\beta'+2r\alpha''-3r\beta''}{2r} \; .
\label{sistema_Gvarphivarphi_3}
\end{equation}
When we use Eqs. (\ref{solucao_phi}) and (\ref{solucao_beta}) in Eq. (\ref{sistema_Gvarphivarphi_3}), the solution for $\alpha$ is
\begin{equation}
\alpha = C_4 + \frac{1}{2} \left [ \ln r - k_1 \ln r + \ln \left ( C_3 + r^{2 k_1} \right ) \right ] \; ,
\label{solucao_alpha}
\end{equation}
where $C_3$, $C_4$ and $k_1$ are constants, with
\begin{equation}
k_1=\sqrt{1-4 C_1^{\; 2}-4 C_a} \; \geq \; 0 \; .
\label{k1}
\end{equation}
Now, when we use the solutions (\ref{solucao_phi}), (\ref{solucao_beta}) and (\ref{solucao_alpha}) in Eq. (\ref{Gtt+Grr_2}), the solutions for $M$ are
\begin{equation}
M(r)_{\pm} = C_5 \; \sqrt{ r^{1 \; - \; k_1} \; \left( C_3 +
r^{2 k_1} \right) } \; \pm \; \sqrt {-\; \frac {r^{1 + 3 k_1}} {C_3  \; \E^{2 C_4} \; \left ( C_3 + r^{2 k_1} \right) }}
\; ,
\label{solucao_M}
\end{equation}
where $C_5$ is a constant. Summarizing, the exact solutions for the system (\ref{sistema_Gtt})-(\ref{sistema_Boxphi}) read
\begin{align}
\phi(r)&=C_2+C_1 \ln r \; , \tag{\ref{solucao_phi}} \\
\beta(r) &= \alpha(r)-\left( C_b+C_a \ln r \right) \; , \tag{\ref{solucao_beta}}\\
\alpha(r) &= C_4 + \frac{1}{2} \left [ \ln r - k_1 \ln r + \ln \left ( C_3 + r^{2 k_1} \right ) \right ] \; , \tag{\ref{solucao_alpha}} \\
M(r)_{\pm} &= C_5 \; \sqrt{ r^{1 \; - \; k_1} \; \left( C_3 +r^{2 k_1} \right) } \;
\pm \; \sqrt {-\; \frac {r^{1 + 3 k_1}} {C_3  \; \E^{2 C_4} \;
\left ( C_3 + r^{2 k_1} \right) }} \; , \tag{\ref{solucao_M}}
\end{align}
where $r>0$, $\frac{\beta(r)}{\alpha(r)} = C_{\alpha\beta} \approx 2$ (e.g., the nearly cylindrical symmetry condition) and, for the solutions to be real, it is necessary that
\begin{align}
%k_1=\sqrt{1-4 C_1^{\; 2}-4 C_a} & \; \geq \; 0 \; , \tag{\ref{k1}} \\
C_3 & \; < \;  0 \; , \label{C3_negativo} \\
|C_3| & \; < \; r^{2 k_1} \; . \label{condicao_C3}
\end{align}
According with the Eqs. (\ref{C3_negativo}) and (\ref{condicao_C3}), $C_3$ can be written as
\begin{equation}
C_3=-\;a_{C_3}\; r_{min}^{\, 2 k_1} \; ,
\label{C3}
\end{equation}
where $0<a_{C_3}<1$ and $r_{min}=\min r$, a value for the radial coordinate below which the system
cannot be associated to the vacuum.

A final remark on  recovering GR out from this scalar-tensorial formalism:  As we are working with vacuum dynamical equations we cannot implement the condition $\omega\rightarrow \infty$ in order to keep track of the GR limit. However, it is possible equivalently to suppress the scalar field dynamics taking $C_1=0$. Henceforward, we shall be making contact with GR by means of this ($C_1=0$) necessary condition.

\section{Spacetime properties}
\label{Secao3}

%\subsection{Kretschmann scalars}

From the line element (\ref{metrica}), with the solutions (\ref{solucao_beta}), (\ref{solucao_alpha}) and (\ref{solucao_M}), the Kretschmann scalars $R^2$, $R_{\mu\nu}R^{\mu\nu}$, $R_{\alpha\beta\mu\nu}R^{\alpha\beta\mu\nu}$ and $C_{\alpha\beta\mu\nu}C^{\alpha\beta\mu\nu}$, where $C_{\alpha\beta\mu\nu}$ stands for the Weyl tensor, in the Einstein frame, are functions of $\alpha$, $\beta$, $M$ and their first and second derivatives with respect to $r$.
In the Jordan-Fierz frame, the line element is rewritten as (\ref{variaveis_Jordan-Einstein_1}) and, with the same solutions (\ref{solucao_beta}), (\ref{solucao_alpha}) and (\ref{solucao_M}) plus the solution (\ref{solucao_phi}), the Kretschmann scalars are functions of $\alpha$, $\beta$, $M$, $\phi$ and their first and second derivatives with respect to $r$, and of the parameter $\omega$.
In the two frames, the results show that the spacetime is well behaved  for $r>0$.

\begin{figure}[h!]
\begin{center}
\includegraphics[width=12cm]{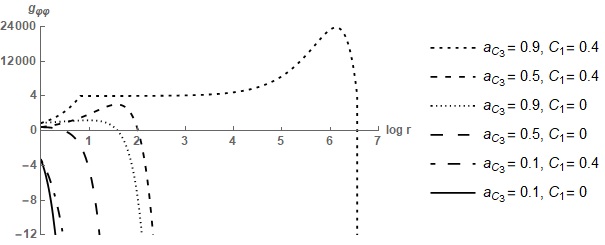}
\caption{Behavior of $g_{\varphi\varphi}$ for distinct $a_{C_3}$ and $C_1$ values, with $C_4=0.1$, $C_5=1$, $C_a=0.08$, $r_{min}=1$ and $M=M_-$.
%When $C_1=0$, the scalar field $\phi$ is constant and the gravitation is the GR; when $C_1=0.4$, we see what happens by %the BD gravitation. It is possible to note that, for a same $a_{C_3}$, the region where there is no CTC (when %$g_{\varphi\varphi}>0$) is larger when the gravitation is that of BD.
When $g_{\varphi\varphi}>4$ the adopted scale for the vertical axis is different from that when $g_{\varphi\varphi} \leq 4$.
}
\label{CTC_BD_X_RG_grafico}
\end{center}
\end{figure}

When the $\varphi$ coefficient of the line element (\ref{metrica}) is negative, CTC's appear. For this, it is
necessary that the following  inequality must hold
\begin{equation}
g_{\varphi\varphi} <  0 \; .
\label{condicao_CTC}
\end{equation}
Then, considering that
\begin{equation}
g_{\varphi\varphi}=\E^{-2 \alpha} \; r^2 - M^2
\label{gvarphivarphi}
\end{equation}
and with the choice of certain constants for the solutions (\ref{solucao_alpha}) and (\ref{solucao_M}), it is possible to determine the intervals for the coordinate $r$ where CTC's appear. For a specific example and considering
adimensional parameters, when $C_4=0.1$, $C_5=1$, $C_a=0.08$, $r_{min}=1$ and the $M$ function is taken with the negative sign in the second term of the Eq. (\ref{solucao_M}), i.e., $M=M_-$, Figure \ref{CTC_BD_X_RG_grafico} shows the behavior of $g_{\varphi\varphi}$ for distinct $a_{C_3}$ and $C_1$ values. As already remarked in Section \ref{Secao2},
if $C_1=0$ the scalar field $\phi$ is constant and the theory approaches GR.

Thus, we can observe in Figure \ref{CTC_BD_X_RG_grafico} that, for a same $a_{C_3}$, the region where there
is no CTC (when $g_{\varphi\varphi}>0$) is larger when we are in the context of BD theory. This reasoning does not apply to $a_{C_3}=0.1$ because, in this case, there is CTC in both theories for any value of the coordinate $r$. For $a_{C_3}=0.9$, there is no CTC up until to the region between $r=10^6$ and $r=10^7$, for the BD gravitation, and between $r=10^1$ and $r=10^2$, for the GR gravitation. When $a_{C_3}=0.5$, in BD, that region is between
$r=10^2$ and $r=10^3$, whereas, in GR, that region is between $r=10^0$ and $r=10^1$.
The fact that the region where there are CTC's, defined from a certain distance from the $z$-axis (and not up to a certain distance, as with the straight spinning cosmic strings \cite{Jensen1992, artigo_corda_reta}) resembles what
is predicted in the Gödel solution, when the CTC's appear when the source distance is from a so-called \textit{critical radius} \cite{Godel1949}.

\section{Rotational curves in galaxies}
\label{Secao4}

The Lagrangian of a particle in the spacetime described by the metric (\ref{metrica}) is given by \cite{Matos2014}
\begin{equation}
2 \mathcal{L}=
-\;\E^{2 \alpha}\dot{t}^2 - M^2 \dot{\varphi}^2 -
 2 \E^\alpha M \dot{t} \dot{\varphi} +
 r^2 \E^{-2 \alpha} \dot{\varphi}^2 +
 \E^{2 \left( \beta - \alpha\right)} \dot{r}^2 +
 \E^{2 \left(\beta - \alpha\right)} \dot{z}^2 \; .
\label{Lagrangiana}
\end{equation}
The canonical momenta associated with the Lagrangian (\ref{Lagrangiana})
%$p_{x^\mu}=\frac {\partial\mathcal{L}}{\partial\dot{x}^\mu}$
are
\begin{eqnarray}
p_t &=&\frac {\partial\mathcal{L}}{\partial\dot{t}}=-E \; , \label{pt} \\
p_\varphi &=&\frac {\partial\mathcal{L}}{\partial\dot{\varphi}}=L \; , \label{pvarphi} \\
p_r &=&\frac {\partial\mathcal{L}}{\partial\dot{r}} \; , \label{pr} \\
p_z &=&\frac {\partial\mathcal{L}}{\partial\dot{z}} \; , \label{pz}
\label{momentos}
\end{eqnarray}
where $E$ and $L$ are conserved quantities \cite{Matos2014}.
By Eqs. (\ref{pt}) and (\ref{pvarphi}), we have
\begin{eqnarray}
\dot{t} &=& -\;\frac{\E^{-2\alpha}\left( -Er^2+L\E^{3\alpha}M+E\E^{2\alpha}M^2 \right)}{r^2} \; , \label{tponto} \\
\dot{\varphi} &=& \frac{\E^\alpha\left( L\E^\alpha+EM \right)}{r^2} \; . \label{varphiponto}
\end{eqnarray}
The Hamiltonian of this particle is
\begin{equation}
\mathcal{H}=p_\mu \dot{x}^\mu-\mathcal{L}=p_t \dot{t}+p_\varphi \dot{\varphi}+p_r \dot{r}+p_z \dot{z}-\mathcal{L} \; .
\label{Hamiltoniana}
\end{equation}
Let us consider the stars as truly test particles moving in a circular motion in the equatorial plane around the center of a galaxy.
Thus, $\dot{r}=0$ and $\dot{z}=0$. Let us further consider that the Hamiltonian is normalized to be equal to $-\;\frac{1}{2}$ for simplicity\footnote{Notice that any non-vanishing normalization may be chosen.} \cite{Matos2014}.
Hence, the system composed by the equations
\begin{eqnarray}
\mathcal{H}+\frac 1 2 &=& 0 \label{Sistema_H_+_partial_H_eq_1} \; ,\\
\frac \partial {\partial r} \left( \mathcal{H}+\frac 1 2 \right) &=& 0 ,
\label{Sistema_H_+_partial_H_eq_2}
\end{eqnarray}
allows solutions to be obtained for the energy $E$ and the moment $L$:
\begin{eqnarray}
E &=& E\left( r, \alpha, M, \alpha', M' \right) \; , \label{solucao_E} \\
L &=& L\left( r, \alpha, M, \alpha', M' \right) \; . \label{solucao_L}
\end{eqnarray}
The solutions (\ref{solucao_E}) and (\ref{solucao_L}) applied in Eqs. (\ref{tponto}) and (\ref{varphiponto}) provide the angular velocity
\begin{eqnarray}
\Omega_{\pm}&=&\frac {d\varphi}{dt}=\frac{\frac{d\varphi}{d\tau}}{\frac{dt}{d\tau}}=\frac{\dot\varphi}{\dot t} \nonumber \\
&=& -\;\frac{2\E^{2\alpha}\alpha'}{\E^\alpha \left( M'+M\alpha' \right)\pm \sqrt{-4r\alpha'\left( -1+r\alpha' \right)+\E^{2\alpha}\left( M'-M\alpha' \right)^2}} \; .
\label{velocidade_angular}
\end{eqnarray}
With the metric (\ref{metrica}) and considering the prescription established by Chandrasekhar $d\tau^2=-ds^2$ \cite{Chandrasekhar1983}, we have
\begin{eqnarray}
d\tau^2 & = & \frac{\E^{2\alpha}r^2}{-e^{2\alpha}M^2+r^2}\;dt^2 \left\{ 1-\frac{\E^{-4\alpha+2\beta}\left( -\E^{2\alpha}M^2+r^2\right)}{r^2} \left [ \left( \frac{dr}{dt} \right)^2+ \left( \frac{dz}{dt} \right)^2 \right ] \right. \nonumber \\
& - & \left. \; \frac{\E^{-4\alpha}\left( \E^{2\alpha}M^{2}-r^2 \right)^2 \left( \frac{d\varphi}{dt} - \frac{\E^{3\alpha}M}{-\E^{2\alpha}M^{2}+r^2} \right)^2}{r^2} \right\} \; .
\label{dtau2}
\end{eqnarray}
Considering $u^0=\frac{dt}{d\tau}$, the Eq. (\ref{dtau2}) can be rewritten as
\begin{equation}
1=\frac{\E^{2\alpha}r^2}{-\E^{2\alpha}M^2+r^2} \left( u^0 \right)^2 \left( 1-v^2 \right) \; ,
\end{equation}
where
\begin{equation}
v^2=v_r^{\; 2}+v_z^{\; 2} + v_\varphi^{\; 2} \; .
\end{equation}
Therefore, the ratio between the tangential velocity $v_\varphi$ and the velocity of light $c$ can be written as
\begin{equation}
{ \left( \frac{v_{\varphi}}{c} \right) }_{\Omega_{\pm}M_{\pm}}=
\sqrt
{\frac{e^{-4\alpha}\left( e^{2\alpha}M_\pm^{\;2}-r^2 \right)^2 \left( \Omega_\pm - \frac{e^{3\alpha}M_\pm}{-e^{2\alpha}M_\pm^{\;2}+r^2} \right)^2}{r^2}} \; .
\label{velocidade_tangencial}
\end{equation}

\begin{figure}[h!]
\begin{center}
\includegraphics[width=12cm]{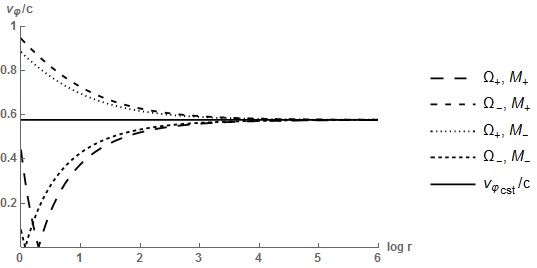}
\caption{When $r_{min}=1$, $k_1=0.5$, $a_{C_3}=0.5$, $C_4=0.1$ and $C_5=-10$, different combinations of the $\Omega$ and $M$ functions in the Eq. (\ref{velocidade_tangencial}) promote distinct tangential velocity functions. It is interesting to note that all functions tend to have the same constant velocity ${v_{\varphi}}_{cst}$.
A similar behavior can be seen for when $\Omega=\Omega_+$ and $M=M_+$ or $\Omega=\Omega_-$ and $M=M_-$ in relation to the velocities of stars in galaxies of work \cite{Rubin1980}, which tend to a constant value in the
approximate range of three orders of magnitude.}
\label{v_varphi_geral_grafico}
\end{center}
\end{figure}

Figure \ref{v_varphi_geral_grafico} shows the ratio $\frac{v_\varphi}{c}$ for different combinations between the functions $\Omega$ and $M$, as predicted by the Eq. (\ref{velocidade_tangencial}), when $r_{min}=1$, $k_1=0.5$, $a_{C_3}=0.5$, $C_4=0.1$ and $C_5=-10$.
It is important to notice that as $r$ increases the velocities tend to a single constant value
\begin{equation}\label{vvarphicstc}
\frac{{v_\varphi}_{cst}}{c}= \sqrt{\frac{1-k_1}{1+k_1}} \; ,
\end{equation}
which is determined when we compute the limit of the ratio $\frac{v\varphi}{c}$ for large $r$. This limit is represented in Figure \ref{v_varphi_geral_grafico} by a solid line. The behavior of the tangential velocities
for the cases where $\Omega=\Omega_+$ and $M=M_+$ or $\Omega=\Omega_-$ and $M=M_-$
is similar to that observed in relation to the velocities of the stars farthest from the centers of the 21 Sc type galaxies studied in Ref. \cite{Rubin1980}.
In Ref. \cite{Rubin1980}, the velocity growth is noticed as the distances from the stars to the centers of the galaxies grow by approximately 2 to 3 orders of magnitude, from $0.5$ kpc to $82.3$ kpc, until they begin to tend to a constant value. This is similar to what is observed in Figure \ref{v_varphi_geral_grafico}, when the growth and consequent tendency for a constant velocity occurs in the range of about $10^0$-$10^3$ m.
This similarity of behaviors was the main motivation to determine if there would be a function (\ref{velocidade_tangencial}) that would fit what was observed in Ref. \cite{Rubin1980}.

For the necessary fit, we considered $C_4=1$ and the choice for the ranges of the constants $a_{C_3}$, $r_{min}$ and the constant velocity ${v_{\varphi}}_{cst}$ from the Eq. (\ref{vvarphicstc}) was performed between those defined by Eqs. (\ref{aC3})-(\ref{velocidade_tangencial_constante}):
\begin{eqnarray}
  a_{C_3} &=& 0.001 n \; , \;\;\; n=1,2,\ldots,999 \; , \label{aC3} \\
  r_{min} &=& 10^{13.0\;+\;0.1 n} \; , \;\;\; n=0,1,\ldots,70 \; , \label{rmin} \\
  {v_\varphi}_{cst}&=& \min {v_\varphi}_{msr} + n \; , \;\;\; n=0,1,\ldots,\max {v_\varphi}_{msr}- \min {v_\varphi}_{msr} \; , \label{velocidade_tangencial_constante}
\end{eqnarray}
where ${v_\varphi}_{msr}$ are the measured velocities of the galaxy under study \cite{Rubin1980}.

The coefficient of determination, $R^2$, is defined by the Eq. (\ref{R2}):
\begin{equation}
\label{R2}
R^2= 1- \; \frac{\sum{\left( {v_\varphi}_{msr}-{v}_\varphi \right)^2}}{\sum {v_\varphi}_{msr}^{\, 2}}\; ,
\end{equation}
where ${v}_\varphi$ is the calculated velocity by the Eq. (\ref{velocidade_tangencial}), when $\Omega=\Omega_+$ and $M=M_+$. Although the coefficient of determination $R^2$ is not the most suitable for testing a non linear regression model,
it is adopted here as a standard definition for the values of the constants $a_{C_3}$, $r_{min}$ and ${v_\varphi}_{cst}$.
The values chosen are those that result in the highest coefficient of determination with the
uncorrected total sum of squares.

As for the constant $C_5$, it is related to $a_{C_3}$, $r_{min}$ and ${v_\varphi}_{cst}$, and has been defined in such a way that CTC does not occur in the range of
$0.5$-$82.3$ kpc. For this to happen, the function $M=M_+$, Eq. (\ref{solucao_M}), was rewritten as
\begin{equation}
M(r)_+=C_5 M_1(r) + M_2(r) \; ,
\label{solucao_M_alternativa}
\end{equation}
where
\begin{eqnarray}
M_1(r) &=& \sqrt{ r^{1 \; - \; k_1} \; \left( C_3 +
r^{2 k_1} \right) } \; , \\
\label{M1}
M_2(r) &=& \sqrt {-\; \frac {r^{1 + 3 k_1}} {C_3  \; \E^{2 C_4} \; \left ( C_3 + r^{2 k_1} \right) }} \; .
\label{M2}
\end{eqnarray}
Thus,
\begin{equation}
{C_5}_{g_{\varphi\varphi}>0} =
%f\left( a_{C_3}, r_{min}, {v_\varphi}_{cst} \right) =
-\; \frac {M_2\left( r=a_{C_5}r_{min}\right)} {M_1\left( r=a_{C_5}r_{min}\right)} \; ,
\label{C5}
\end{equation}
with $a_{C_5}=10^6$.
When the best fit of the function (\ref{velocidade_tangencial}) is sought for the 21 galaxies of Ref. \cite{Rubin1980}, if the constant $C_5$ is calculated according to the Eq. (\ref{C5}), the first term of the Eq. (\ref{gvarphivarphi}) is always larger than the second term, and, therefore, there will be no CTC, because $g_{\varphi\varphi}>0$.
However, the $C_5$ definition leads to $v_\varphi \geq {v_\varphi}_{cst}$ in this case, which runs counter to the observations, since we expect $v_\varphi \leq {v_\varphi}_{cst}$.
For this problem to be solved, simply multiply $C_5$ by $-1$, that is, use the same constant $C_5$ determined before, but with a positive sign.
The tangential velocity will have adequate values, but this time CTC's cannot be precluded, since that $g_{\varphi\varphi}<0$.
This happens because the values of $C_5$ for $v_\varphi \leq {v_\varphi}_{cst}$ are out of the range of $C_5$ where CTC does not appear.
For example, the function (\ref{velocidade_tangencial}) has its best fit to the measurements of the NGC 1087 galaxy when $\Omega=\Omega_+$, $M=M_+$, $C_4=1$, ${a_C}_3=0.723$, $r_{min}=10^{16.0}$ m, ${v_\varphi}_{cst}=130$ km/s and
\begin{eqnarray}
{C_5}_{v_\varphi \leq {v_\varphi}_{cst}} &=& +\;\frac {M_2\left( r=a_{C_5}r_{min}\right)} {M_1\left( r=a_{C_5}r_{min}\right)} \label{C5comCTC} \\
&=& 4.3266\cdot 10^{-17} \; . \nonumber
\end{eqnarray}
With this fitting CTC occurs, because in order to have $g_{\varphi\varphi}>0$ it is necessary that
\begin{equation}\label{}
-4.3267\cdot 10^{-17}<{C_5}_{g_{\varphi\varphi}>0}<-4.3265\cdot 10^{-17} \;, \nonumber
\end{equation}
but this value leads to $v_\varphi \geq {v_\varphi}_{cst}$.

To sum up, Eqs. (\ref{C5}) and (\ref{C5comCTC}) allow us to write the following relation
\begin{equation}
\label{relacaoC5}
{C_5}_{g_{\varphi\varphi}>0} = -\;{C_5}_{v_\varphi \leq {v_\varphi}_{cst}} \; .
\end{equation}
If we obey Eq. (\ref{relacaoC5}), when
\begin{equation}
\left | {C_5}_{g_{\varphi\varphi}>0} \right |=\left | {C_5}_{v_\varphi \leq {v_\varphi}_{cst}} \right | \; , \nonumber
\end{equation}
if $C_5<0$, there is no CTC, but $v_\varphi \geq {v_\varphi}_{cst}$; on the other hand, if $C_5>0$, $v_\varphi \leq {v_\varphi}_{cst}$, but there is CTC.

\begin{figure}
\begin{center}
\includegraphics[width=12cm]{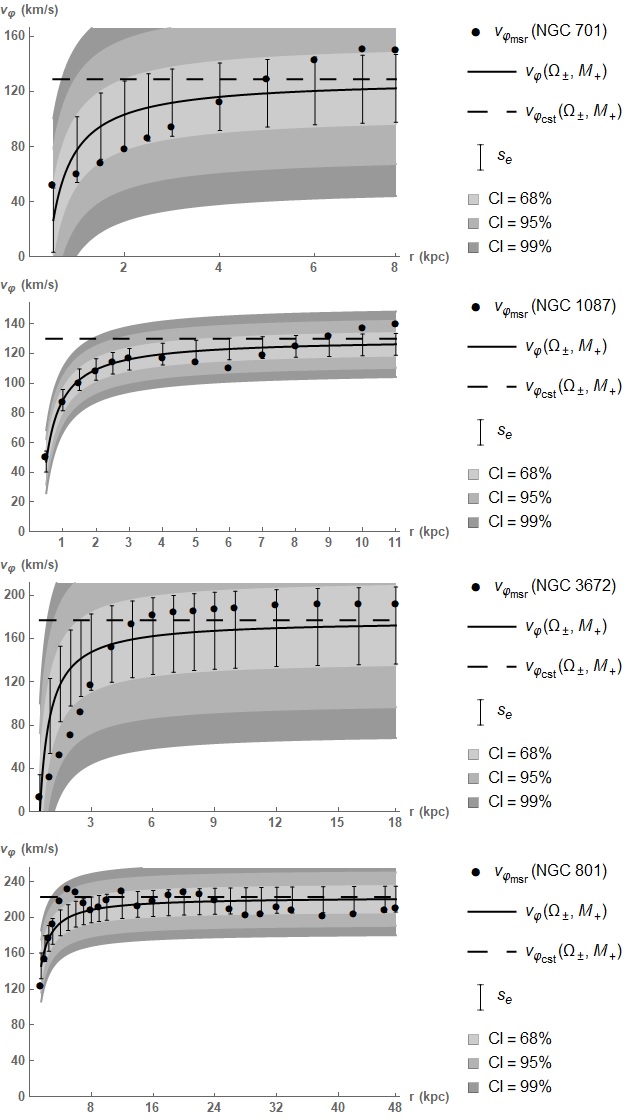}
\caption{Stars velocities of the NGC 701, NGC 1087, NGC 3672 and NGC 801 galaxies \cite{Rubin1980} compared to the best fit of the function (\ref{velocidade_tangencial}).
When $\Omega=\Omega_+$, the curves show the fits for when the $C_4$ and $C_5$ constants are real and there is CTC in the range represented in the graph, because $g_{\varphi\varphi}<0$; when $\Omega=\Omega_-$, $C_4$ and $C_5$ are imaginary, $g_{\varphi\varphi}>0$, and there is no CTC, and the metric is Euclidean, because $g_{tt}>0$. The values of the constants and the coefficient of determination for each galaxy are in the Table \ref{4Galaxias_C4RiC5Ri_tabela}, and the values of the measured and calculated velocities, the standard errors and the CI's are in the Tables \ref{NGC_701_CI}-\ref{NGC_801_CI} of the Appendix \ref{apendice}.}
\label{4Galaxias_grafico}
\end{center}
\end{figure}

Figure \ref{4Galaxias_grafico} shows the behavior of the fitted function (\ref{velocidade_tangencial}) for the NGC 701, NGC 1087, NGC 3672 and NGC 801 galaxies, when $C_5>0$, that is, when $v_\varphi$ takes suitable values with the observations, but $g_{\varphi\varphi}<0$.

The confidence intervals (CI's) of the velocities were calculated according to Eq. (\ref{vvarphiCI}):
\begin{equation}\label{vvarphiCI}
{v_\varphi}_{CI}\left( r \right)=v_\varphi\left( r \right) \, \pm \, t\left( n-p, 1-\frac{1-CI}{2}\right) \,s_e\left( r \right) \; ,
\end{equation}
where $t$ is the $1-\frac{1-CI}{2}$ quantile of the Student t-distribution with $n-p$ degrees of freedom, being $n$ the number of measurements taken for the galaxy and $p=1$, since all constants of the function (\ref{velocidade_tangencial}) have been previously defined. Besides, the standard error $s_e$ is
\begin{equation}\label{se}
s_e\left( r \right)=\sqrt{s^2\left( 1+\frac{v_\varphi\left( r \right)^{2}}{\sum{v_\varphi^{\,2}}}\right)} \; ,
\end{equation}
with the estimated error variance $s^2$ equal to
\begin{equation}\label{variance}
s^2 = \frac{\sum{\left( {v_\varphi}_{msr}-{v}_\varphi \right)^2}}{n-p} \; .
\end{equation}
%and $V_\varphi$ is the matrix
%V_\varphi &=&
%\begin{bmatrix}
%  {v_\varphi}\left( r_1 \right) \\
%  {v_\varphi}\left( r_2 \right) \\
%  \vdots \\
%  {v_\varphi}\left( r_n \right)
%\end{bmatrix} \; ,
%which contains all the calculated velocities $v_\varphi$ for the $n$ radii $r$ of the measured velocities %${v_\varphi}_{msr}$.
CI's equal to 68$\%$, 95$\%$ and 99\% for the model applied in the four galaxies are represented in the Figure \ref{4Galaxias_grafico}. The values of the velocities ${v_\varphi}_{msr}$ and $v_\varphi$, the standard errors $s_e$, and the CI's shown in the same Figure are in the Tables \ref{NGC_701_CI}-\ref{NGC_801_CI} of the Appendix \ref{apendice}.
In Table \ref{4Galaxias_C4RiC5Ri_tabela} we see the values of the constants with which the highest coefficient of determination $R^2$ was obtained, providing the best fit of the function (\ref{velocidade_tangencial}) for each one of the four galaxies at hand.
Among the 21 studied galaxies, NGC 701 and NGC 3672 are the ones with the lowest coefficients of determination, and NGC 1087 and NGC 801 are those with the highest coefficients. These two last galaxies have smaller standard errors than the two first, ensuring a better accuracy of the calculated velocities.

\begin{table}[h!]
\caption{Values of the components $g_{tt}$ and $g_{\varphi\varphi}$, the constants and the determination coefficients for the fits of the function (\ref{velocidade_tangencial}) made for NGC 701, NGC 1087, NGC 3672 and NGC 801 galaxies, shown in Figure \ref{4Galaxias_grafico}.
$R_{farthest}$ is the radial distance of outermost measured velocity.
When the constant $C_5$ is real, it is defined by Eq. (\ref{C5comCTC}).}
\begin{center}
%\centering
\begin{tabular}{r | c | c | c | c | c | c | c | c}
%\hline\hline
& \multicolumn{2}{c |}{NGC 701} & \multicolumn{2}{c |}{NGC 1087} & \multicolumn{2}{c |}{NGC 3672} & \multicolumn{2}{c}{NGC 801} \\
\hline\hline
\; $R_{farthest}$ (kpc) \; & \multicolumn{2}{c |}{7.7} & \multicolumn{2}{c |}{11.0} & \multicolumn{2}{c |}{17.8} & \multicolumn{2}{c}{47.4} \\ \hline
$\Omega$ \; & $\Omega_+$ & $\Omega_-$ & $\Omega_+$ & $\Omega_-$ & $\Omega_+$ & $\Omega_-$ & $\Omega_+$ & $\Omega_-$ \\ \hline
$M$ \; & \multicolumn{8}{c}{$M_+$} \\ \hline
${v_\varphi}_{cst}$ (km/s) \; & \multicolumn{2}{c |}{129} & \multicolumn{2}{c |}{130} & \multicolumn{2}{c |}{177} & \multicolumn{2}{c}{223} \\ \hline
$\log \; r_{min}$ \; & 16.1 & 15.8 & 16.0 & 15.8 & 16.5 & 16.1 & 16.5 & 16.3 \\ \hline
$a_{C_3}$ \; & \; 0.699 \; & \; 0.696 \; & \; 0.723 \; & \; 0.454 \; & \; 0.331 \; & \; 0.522 \; & \; 0.562 \; & \; 0.353 \; \\ \hline
$C_4$ \; & 1 & $\frac{\pi}{2}\I$ & 1 & $\frac{\pi}{2}\I$ & 1 & $\frac{\pi}{2}\I$ & 1 & $\frac{\pi}{2}\I$ \\ \hline
$C_5$ \; & $\frac{M_2}{M_1}$ & $\I$ & $\frac{M_2}{M_1}$ & $\I$ & $\frac{M_2}{M_1}$ & $\I$ & $\frac{M_2}{M_1}$ & $\I$ \\ \hline
$R^2$ \; & \multicolumn{2}{c |}{0.958} & \multicolumn{2}{c |}{0.996} & \multicolumn{2}{c |}{0.953} & \multicolumn{2}{c}{0.996} \\ \hline
$g_{\varphi\varphi}$ \; & \multirow{2}{*}{$<0$} & \multirow{2}{*}{$>0$} & \multirow{2}{*}{$<0$} & \multirow{2}{*}{$>0$} & \multirow{2}{*}{$<0$} & \multirow{2}{*}{$>0$} & \multirow{2}{*}{$<0$} & \multirow{2}{*}{$>0$} \\ \cline{1-1}
$g_{tt}$ \; & & & & & & &
\end{tabular}
\label{4Galaxias_C4RiC5Ri_tabela}
\end{center}
\end{table}

There is an alternative to the function (\ref{velocidade_tangencial}) in order to suitably fit
velocities with the observations and also not allow the existence of CTC's.
But this implies the need of both $\E^{-C_4}$ and $C_5$ to assume pure imaginary values, which will make the
solution (\ref{solucao_M}) be  pure imaginary number and the solutions (\ref{solucao_beta}) and (\ref{solucao_alpha}) become complex, but the component $g_{\varphi\varphi}$ (\ref{gvarphivarphi}) and the tangential velocity (\ref{velocidade_tangencial}) remain real, as well as the metric (\ref{metrica}).
So, if we now consider $\Omega=\Omega_-$ instead of $\Omega=\Omega_+$, and, for example, $C_4=\frac{\pi}{2}\I$ and $C_5=\I$, $g_{\varphi\varphi}$ will be positive and the tangential velocity will
assume values of the expected order of magnitude.
The results for the best fits of the NGC 701, NGC 1087, NGC 3672 and NGC 801 galaxies for this last alternative
can also be seen in Figure \ref{4Galaxias_grafico} and in the Table \ref{4Galaxias_C4RiC5Ri_tabela}.
%Momento Angular Complexo.

%\begin{figure}[h!]

As for the metric (\ref{metrica}), when $C_4=\frac{\pi}{2}\I$, the component $g_{tt}$ reads
\begin{equation}\label{gttpositiva}
g_{tt}=r^{1-k_1} \left( C_3+r^{2 k_1}\right) \; .
\end{equation}
For the fit of the function (\ref{velocidade_tangencial}),
the conditions (\ref{C3_negativo}) and (\ref{condicao_C3}) were maintained for the metric to be real. Thus, $C_3+r^{2 k_1}>0$, which leads to $g_{tt}>0$.
Being still
$g_{\varphi\varphi}>0$ and the other spatial components $g_{rr}$ and $g_{zz}$ also positive, the metric (\ref{metrica}) becomes an Euclidean metric.

\section{The ratio $\frac {\beta}{\alpha}$: An alternative to explain the rotational curves?}
\label{Secao5}

As we make use of a nearly cylindrical symmetry, it is necessary that the ratio between the functions (\ref{solucao_beta}) and (\ref{solucao_alpha}) gives $\frac{\beta}{\alpha}= C_{\alpha\beta} \approx 2$. Since we
are interested in the behavior of $v_\varphi$ for the $r = 0,5$-$82.3$ kpc range, it is important to calculate the limit for large $r$ of the $\frac{\beta}{\alpha}$ ratio. This limit is constant and can be written as
\begin{equation}
{C_{\alpha\beta}}_{cst} = 1-\;\frac {2 C_a}{1+k_1} \; ,
\end{equation}
leading to
\begin{equation}
C_a = -\;\frac 1 2 \left( {C_{\alpha\beta}}_{cst}-1 \right) \left( 1+k_1 \right) \; .
\label{Ca}
\end{equation}
According to Eq. (\ref{Ca}), when ${C_{\alpha\beta}}_{cst}=2.002$ (that is, 1$\%$ bigger than 2) and, for example, ${v_\varphi}_{cst}=200$ km/s (this value has the suitable order of magnitude, because, for all $21$ studied galaxies, $96\; \textrm{km/s} \leq {v_\varphi}_{cst} \leq 266\; \textrm{km/s}$), with $c=3\cdot 10^5$ km/s and
\begin{equation}
\frac{{v_\varphi}_{cst}}{c}= \sqrt{\frac{1-k_1}{1+k_1}} \; , \tag{\ref{vvarphicstc}}
\end{equation}
the constant $C_a$ equals $-1.002$.
Thus, considering
\begin{equation}
k_1=\sqrt{1-4 C_1^{\; 2}-4 C_a} \; \geq \; 0 \tag{\ref{k1}}
\end{equation}
we have $C_1=\pm 1.001\neq 0$, that is, the scalar field $\phi$ is not constant (see the Eq. (\ref{solucao_phi})), which indicates that the behavior of tangential velocities of the stars seen so far can not be described by GR, and the use of BD gravitation is indeed required.

If the modeling option is in the framework of the GR gravity, i.e., with $C_1=0$, by the Eqs. (\ref{k1}) and (\ref{Ca}), with ${C_{\alpha\beta}}_{cst}=2.002$, we have
\begin{eqnarray}
C_a &=& \frac {1-k_1^{\;2}}{4} \; ,
\label{CaC1nulo} \\
C_a &=& -\;\frac 1 2 \left( 2.002-1 \right) \left( 1+k_1 \right) \; .
\label{CaCalphabetacst2002}
\end{eqnarray}
By the Eq. (\ref{k1}), $k_1\geq 0$.
So, the only valid solution for the system with the Eqs. (\ref{CaC1nulo}) and (\ref{CaCalphabetacst2002}) is $k_1=3.004$.
This last result leads to ${v_\varphi}_{cst}=0.707c\I$, which is not possible, because ${v_\varphi}_{cst}$ must be real.
%and, for all $21$ studied galaxies, $96\; \textrm{km/s} \leq {v_\varphi}_{cst} \leq 266\; \textrm{km/s}$, in addition to the fact that $\frac {{v_\varphi}_{cst}} c\leq 1$ \cite{Chandrasekhar1983}.

%Therefore, the proposed modeling in cylindrical coordinates requires that the gravitation adopted be that of BD, because the RG, when the equations (\ref{Ca}) and (\ref{CaC1nulo}), with ${v_\varphi}_{cst}$$=$$200$ km/s, leads to $C_{\beta\alpha}=0.999 \ll 2$, which is not suitable, because $C_{\beta\alpha} \approx 2$.

%--------------------------------------------------

\begin{figure}
\begin{center}
\includegraphics[width=16cm]{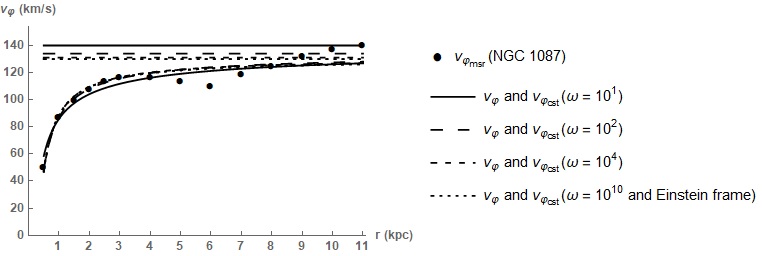}
\caption{Stars velocities of the NGC 1087 galaxy \cite{Rubin1980} compared to the best fits of the function (\ref{velocidade_tangencial}) in the Jordan-Fierz frame.
The values of the constants and the coefficients of determination are in the Table \ref{NGC1087Jordan-Fierz_tabela}.
From $\omega=10^2$, the fit of the function (\ref{velocidade_tangencial}) results in almost identical curves.}
\label{NGC1087Jordan-Fierz_grafico}
\end{center}
\end{figure}

With the line element in the Jordan-Fierz frame (\ref{variaveis_Jordan-Einstein_1}), it is also possible to determine the angular velocity\footnote{Along the calculations we have considered $\kappa$ as well $\omega$ as positive parameters.}:
\begin{equation}\label{velocidade_angular_Jordan}
\Omega_\pm= \pm\;\frac{\E^{5\alpha}\Xi \left( 2\Lambda\alpha ' +\Lambda ' \right)}{\sqrt{\E^{8\alpha}\Lambda^2\Xi^4 - \E^{6\alpha}r\Xi^2 \left( 2\Lambda\alpha'+\Lambda' \right)\left[ 2\Lambda\left( -1+r\alpha' \right) -r\Lambda' \right]}\; \mp \; \E^{4\alpha}\Xi\left( M\alpha'-\Lambda\Xi \right)} \; ,
\end{equation}
where
\begin{eqnarray}
\Lambda &=& \E^{2\kappa \phi} \; ,\\
\Xi &=& M\alpha'-M' \; .
\end{eqnarray}
When we insert the velocity (\ref{velocidade_angular_Jordan}) in the Eq. (\ref{velocidade_tangencial}), we have the ratio $\frac{v_\varphi}{c}$ in the Jordan-Fierz frame, where the parameter $\omega$ now appears explicitly.
In this case, the limit of this ratio for large $r$ tends to the single value
\begin{equation}\label{vvarphicstc_Jordan}
\frac{{v_\varphi}_{cst}}{c}= \sqrt{\frac{\left( 1-k_1^{\,2}\right) \left( 3+2\omega \right)+4C_1\left( C_1+\sqrt{3+2\omega}\,\right)}{\left( 1+k_1 \right)^2 \left( 3+2\omega \right) +4C_1\left[ C_1+\left( 1+k_1\right)\sqrt{3+2\omega}\,\right]}} \; .
\end{equation}
When $\omega\rightarrow\infty$, the limit of this ratio is
\begin{equation}\label{limit_omega_inf_vvarphicstc}
\lim_{\omega\rightarrow\infty} \frac{{v_\varphi}_{cst}}{c}= \sqrt{\frac{1-k_1}{1+k_1}} \; .
\end{equation}
This last result is identical to that one found for the ratio $\frac{{v_\varphi}_{cst}}{c}$ in the Einstein frame (see Eq. (\ref{vvarphicstc})).
This happens because when $\omega\rightarrow\infty$, $\tilde g_{\mu\nu}=g_{\mu\nu}$.
Thus, for a large $\omega$, the behaviors of the velocity $v_\varphi$ are the same in the Einstein and Jordan-Fierz frames.
For example, the Figure \ref{NGC1087Jordan-Fierz_grafico} shows the behavior of the fitted velocity $v_\varphi$ for the NGC 1087 galaxy, with different values of $\omega$.
We consider $C_2=C_4=1$; ${C_{\alpha\beta}}_{cst}=2.002$; ${a_C}_3$, $r_{min}$ and ${v_\varphi}_{cst}$ defined by Eqs. (\ref{aC3})-(\ref{velocidade_tangencial_constante}); $C_5$ defined by Eq. (\ref{C5comCTC}); $C_1$ determined by the system formed by Eqs. (\ref{k1}), (\ref{Ca}) and (\ref{vvarphicstc_Jordan}), depending on the constants $\omega$, ${v_\varphi}_{cst}$ and $C_{\alpha\beta}$.

As $\omega$ grows, the behavior of the velocity $v_\varphi$ is close to that one found in the Einstein frame (also shown in Figure \ref{4Galaxias_grafico}). For instance, when $\omega=10^{10}$, the constants ${a_C}_3$, $r_{min}$ and ${v_\varphi}_{cst}$ coincide in both frames (see the Tables \ref{4Galaxias_C4RiC5Ri_tabela} and \ref{NGC1087Jordan-Fierz_tabela}). It is also interesting to note that, for $\omega=10^2$ or $\omega=10^4$, the coefficient of determination in the Jordan-Fierz is somewhat higher than that found when $\omega=10^{10}$ or in the Einstein frame.

As usual, when using the Jordan-Fierz frame along with a large $\omega$, the BD gravitation must provide (almost) the same results of GR. Nevertheless, the obtained good results with a small $\omega\sim 10^2$ attest that such gravitation is the correct option, reinforcing what was already argued in the first part of this Section. In addition, because of the difference between the fitted curves in the different frames, accurate measurements of the stars velocities may indicate which of the two frames would be most appropriate for the quantitative description of the rotational curves of the galaxies.

\begin{table}[h!]
\caption{Values of the constants and the coefficients of determination for the fits of the function (\ref{velocidade_tangencial}) in the Jordan-Fierz for NGC 1087 galaxy, shown in Figure \ref{NGC1087Jordan-Fierz_grafico}.
The constant $C_1$ was defined by the system of the Eqs. (\ref{k1}), (\ref{Ca}) and (\ref{vvarphicstc_Jordan}), and $C_5$ was defined by the Eq. (\ref{C5comCTC}).
}
\begin{center}
%\centering
\begin{tabular}{c | c | c | c | c | c | c | c | c | c | c | c | c}
%\hline\hline
\multicolumn{13}{c}{NGC 1087} \\
\hline\hline
$\log \; \omega$ & \, $\Omega$ \, & \, $M$ \, & ${v_\varphi}_{cst}$ (km/s) & $\log \; r_{min}$ & \, $a_{C_3}$ \, & \, $C_1$ \, & $C_2$ & $C_4$ & $C_5$ & ${C_{\alpha\beta}}_{cst}$ & \, $R^2$ \, & $g_{\varphi\varphi}$ \\ \hline\hline
\, $1$ \, & \multirow{4}{*}{$\Omega_+$} & \multirow{4}{*}{$M_+$} & \, 140 \, & \, 13.8 \, & 0.697 & \multirow{4}{*}{$f(\omega,{v_\varphi}_{cst},{C_{\alpha\beta}}_{cst})$} & \multicolumn{2}{c |}{\multirow{4}{*}{\, 1 \,}} & \multirow{4}{*}{$\frac{M_2}{M_1}$} & \multirow{4}{*}{2.002} & 0.996 & \multirow{4}{*}{$<0$} \\ \cline{1-1}\cline{4-6}\cline{12-12}
\, $2$ \, & & & \, 134 \, & \, 15.7 \, & 0.324 & & \multicolumn{2}{c |}{} & & & \multirow{2}{*}{0.997} & \\ \cline{1-1}\cline{4-6}
\, $4$ \, & & & \, 131 \, & \, 16.1 \, & 0.387 & & \multicolumn{2}{c |}{} & & & \\ \cline{1-1}\cline{4-6}\cline{12-12}
\, $10$ \, & & & \, 130 \, & \, 16.0 \, & 0.723 & & \multicolumn{2}{c |}{} & & & 0.996 &
\end{tabular}
\label{NGC1087Jordan-Fierz_tabela}
\end{center}
\end{table}

%--------------------------------------------------------------

\section{Final Remarks}
\label{Secao6}

A complete spacetime analysis was performed taking into account a nearly cylindrically symmetry in the context of the BD gravitation. The resulting spacetime is regular and there is a possibility of the existing of CTC's for a certain range of the coordinate $r$.  With an appropriate choice of the integration constants,
it is possible to avoid the CTC's.

The solution presented in this paper is shown to be suitable to generate the rotational curves of the galaxies, because it reproduces the velocity growth behavior in the order of magnitude compatible with the experimental data \cite{Rubin1980}.
For this, the  solution indicates that it is necessary to admit the existence of CTC's in a certain range.
 We notice, however, that this type of possibility is far from consensual. On the one hand the \textit{chronological protection conjecture} says that the laws of physics prohibit the emergence of CTC's \cite{Hawking1992}. On the other hand, interesting works point out that these same laws allow the CTC's and they can appear naturally
 \cite{Thorne1988, Novello1992, Novello1993, Novello1994}.

Returning to the main goal of our paper,  we  present a model in which CTC's are ruled out, but the metric
must be Euclidean. Theories of quantum gravity, which seek to describe phenomena of the time of grand
unification theory (GUT), have indeed made use of such a stratagem \cite{gravitacao_quantica}.

Finally, we would like to stress that the BD gravity adequately describes the observational data as an effective theory of gravity coming from an already unknown completely theory of a quantum gravity.

\begin{acknowledgments}
JMHS would like to thank the CNPq for partial support (grants number 304629/2015-4; 445385/2014-6).
The authors would like to thank R. Floriano da Silva for his collaboration in calculating  the solutions for the metric's functions and O. Grichtchouk for a critical reading of this manuscript.
\end{acknowledgments}

\newpage

\appendix

\section{Measured and calculated velocities, standard errors and confidence intervals for the NGC 701, NGC 1087, NGC 3672 and NGC 801 galaxies}
\label{apendice}

\begin{table}[h!]
\caption{Measured and calculated velocities, standard errors and confidence intervals for the NGC 701 galaxy.}
\begin{center}
%\centering
\begin{tabular}{c | c | c | c | c | c | c }
%\hline\hline
\multicolumn{7}{c}{NGC 701} \\
\hline\hline
\multirow{2}{*}{\; $r$ (kpc) \cite{Rubin1980} \;} & \multirow{2}{*}{\; ${v_\varphi}_{msr}$ (km/s) \cite{Rubin1980} \;} & \multirow{2}{*}{\; ${v_\varphi}$ (km/s) \;} & \multirow{2}{*}{\; $s_e$ (km/s) \;} & \multicolumn{3}{c}{Confidence interval} \\ \cline{5-7}
& & & & $68\%$ & $95\%$ & $99\%$ \\ \hline\hline
0.5 & 52 & 27 & 23 & $ \; [2, 51] \; $ & $ \; [-25, 79] \; $ & $ \; [-47, 101] \; $ \\ \hline
1 & 60 & 78 & 24 & $ \; [53, 103] \; $ & $ \; [25, 131] \; $ & $ \; [2, 153] \; $ \\ \hline
1.5 & 68 & 95 & 24 & $ \; [70, 120] \; $ & $ \; [41, 149] \; $ & $ \; [18, 171] \; $ \\ \hline
2 & 78 & 103 & 24 & $ \; [78, 129] \; $ & $ \; [49, 157] \; $ & $ \; [27, 180] \; $ \\ \hline
2.5 & 86 & 109 & 24 & $ \; [83, 134] \; $ & $ \; [54, 163] \; $ & $ \; [31, 186] \; $ \\ \hline
3 & 94 & 112 & 24 & $ \; [86, 137] \; $ & $ \; [58, 166] \; $ & $ \; [35, 189] \; $ \\ \hline
4 & 112 & 116 & 25 & $ \; [91, 142] \; $ & $ \; [62, 171] \; $ & $ \; [39, 194] \; $ \\ \hline
5 & 129 & 119 & 25 & $ \; [93, 144] \; $ & $ \; [64, 173] \; $ & $ \; [41, 197] \; $ \\ \hline
6 & 143 & 120 & 25 & $ \; [95, 146] \; $ & $ \; [66, 175] \; $ & $ \; [43, 198] \; $ \\ \hline
7 & 151 & 122 & 25 & $ \; [96, 147] \; $ & $ \; [67, 177] \; $ & $ \; [44, 200] \; $ \\ \hline
7.7 & 150 & 122 & 25 & $ \; [97, 148] \; $ & $ \; [67, 177] \; $ & $ \; [44, 200] \; $
\end{tabular}
\label{NGC_701_CI}
\end{center}
\end{table}

\begin{table}[h!]
\caption{Measured and calculated velocities, standard errors and confidence intervals for the NGC 1087 galaxy.}
\begin{center}
%\centering
\begin{tabular}{c | c | c | c | c | c | c }
%\hline\hline
\multicolumn{7}{c}{NGC 1087} \\
\hline\hline
\multirow{2}{*}{\; $r$ (kpc) \cite{Rubin1980} \;} & \multirow{2}{*}{\; ${v_\varphi}_{msr}$ (km/s) \cite{Rubin1980} \;} & \multirow{2}{*}{\; ${v_\varphi}$ (km/s) \;} & \multirow{2}{*}{\; $s_e$ (km/s) \;} & \multicolumn{3}{c}{Confidence interval} \\ \cline{5-7}
& & & & $68\%$ & $95\%$ & $99\%$ \\ \hline\hline
0.5 & 50 & 47 & {7} & $\;[40 , 55]\;$ & $\;[32 , 63]\;$ & $\;[26 , 69]\;$ \\ \cline{1-4}\cline{5-7}
1 & 87 & 89 & 7 & $\;[81 , 96]\;$ & $\;[73 , 104]\;$ & $\;[67 , 110]\;$ \\ \cline{1-4}\cline{5-7}
1.5 & 100 & 102 & 7 & $\;[95 , 110]\;$ & $\;[87 , 118]\;$ & $\;[81 , 124]\;$ \\ \cline{1-4}\cline{5-7}
2 & 108 & 109 & 7 & $\;[102 , 117]\;$ & $\;[94 , 125]\;$ & $\;[88 , 131]\;$ \\ \cline{1-4}\cline{5-7}
2.5 & 114 & 113 & 7 & $\;[106 , 121]\;$ & $\;[98 , 129]\;$ & $\;[92 , 135]\;$ \\ \cline{1-4}\cline{5-7}
3 & 117 & 116 & 7 & $\;[109 , 124]\;$ & $\;[101 , 132]\;$ & $\;[94 , 138]\;$ \\ \cline{1-4}\cline{5-7}
4 & 117 & 120 & 7 & $\;[112 , 127]\;$ & $\;[104 , 135]\;$ & $\;[98 , 142]\;$ \\ \cline{1-4}\cline{5-7}
5 & 114 & 122 & 7 & $\;[114 , 129]\;$ & $\;[106 , 137]\;$ & $\;[100 , 144]\;$ \\ \cline{1-4}\cline{5-7}
6 & 110 & 123 & 7 & $\;[116 , 131]\;$ & $\;[107 , 139]\;$ & $\;[101 , 145]\;$ \\ \cline{1-4}\cline{5-7}
7 & 119 & 124 & 7 & $\;[117 , 132]\;$ & $\;[108 , 140]\;$ & $\;[102 , 146]\;$ \\ \cline{1-4}\cline{5-7}
8 & 125 & 125 & 7 & $\;[117 , 132]\;$ & $\;[109 , 141]\;$ & $\;[103 , 147]\;$ \\ \cline{1-4}\cline{5-7}
9 & 132 & 125 & 7 & $\;[118 , 133]\;$ & $\;[110 , 141]\;$ & $\;[103 , 147]\;$ \\ \cline{1-4}\cline{5-7}
10 & 137 & 126 & 7 & $\;[118 , 133]\;$ & $\;[110 , 142]\;$ & $\;[104 , 148]\;$ \\ \cline{1-4}\cline{5-7}
11 & 140 & 126 & 7 & $\;[119 , 134]\;$ & $\;[110 , 142]\;$ & $\;[104 , 148]\;$
\end{tabular}
\label{NGC_1087_CI}
\end{center}
\end{table}

\begin{table}[h!]
\caption{Measured and calculated velocities, standard errors and confidence intervals for the NGC 3672 galaxy.}
\begin{center}
%\centering
\begin{tabular}{c | c | c | c | c | c | c }
%\hline\hline
\multicolumn{7}{c}{NGC 3672} \\
\hline\hline
\multirow{2}{*}{\; $r$ (kpc) \cite{Rubin1980} \;} & \multirow{2}{*}{\; ${v_\varphi}_{msr}$ (km/s) \cite{Rubin1980} \;} & \multirow{2}{*}{\; ${v_\varphi}$ (km/s) \;} & \multirow{2}{*}{\; $s_e$ (km/s) \;} & \multicolumn{3}{c}{Confidence interval} \\ \cline{5-7}
& & & & $68\%$ & $95\%$ & $99\%$ \\ \hline\hline
0.5 & 14 & 0.02 & 34 & $ \; [-35, 35] \; $ & $ \; [-73, 73] \; $ & $ \; [-100, 100] \; $ \\ \hline
1 & 32 & 89 & 35 & $ \; [53, 124] \; $ & $ \; [15, 162] \; $ & $ \; [-13, 190] \; $ \\ \hline
1.5 & 52 & 118 & 35 & $ \; [82, 154] \; $ & $ \; [44, 192] \; $ & $ \; [16, 220] \; $ \\ \hline
2 & 71 & 133 & 35 & $ \; [97, 169] \; $ & $ \; [58, 207] \; $ & $ \; [30, 235] \; $ \\ \hline
2.5 & 92 & 142 & 35 & $ \; [106, 178] \; $ & $ \; [67, 216] \; $ & $ \; [39, 244] \; $ \\ \hline
3 & 117 & 148 & 35 & $ \; [111, 184] \; $ & $ \; [73, 222] \; $ & $ \; [45, 250] \; $ \\ \hline
4 & 152 & 155 & 35 & $ \; [119, 191] \; $ & $ \; [80, 230] \; $ & $ \; [52, 258] \; $ \\ \hline
5 & 173 & 159 & 35 & $ \; [123, 196] \; $ & $ \; [84, 234] \; $ & $ \; [56, 263] \; $ \\ \hline
6 & 182 & 162 & 35 & $ \; [126, 199] \; $ & $ \; [87, 237] \; $ & $ \; [59, 266] \; $ \\ \hline
7 & 184 & 164 & 35 & $ \; [128, 201] \; $ & $ \; [89, 240] \; $ & $ \; [61, 268] \; $ \\ \hline
8 & 185 & 166 & 35 & $ \; [130, 202] \; $ & $ \; [91, 241] \; $ & $ \; [62, 270] \; $ \\ \hline
9 & 187 & 167 & 36 & $ \; [131, 204] \; $ & $ \; [92, 242] \; $ & $ \; [63, 271] \; $ \\ \hline
10 & 188 & 168 & 36 & $ \; [132, 205] \; $ & $ \; [93, 243] \; $ & $ \; [64, 272] \; $ \\ \hline
12 & 191 & 170 & 36 & $ \; [133, 206] \; $ & $ \; [94, 245] \; $ & $ \; [66, 273] \; $ \\ \hline
14 & 192 & 171 & 36 & $ \; [134, 207] \; $ & $ \; [95, 246] \; $ & $ \; [67, 275] \; $ \\ \hline
16 & 192 & 171 & 36 & $ \; [135, 208] \; $ & $ \; [96, 247] \; $ & $ \; [68, 275] \; $ \\ \hline
17.8 & 192 & 172 & 36 & $ \; [136, 209] \; $ & $ \; [97, 247] \; $ & $ \; [68, 276] \; $
\end{tabular}
\label{NGC_3672_CI}
\end{center}
\end{table}

\begin{table}[h!]
\caption{Measured and calculated velocities, standard errors and confidence intervals for the NGC 801 galaxy.}
\begin{center}
%\centering
\begin{tabular}{c | c | c | c | c | c | c }
%\hline\hline
\multicolumn{7}{c}{NGC 801} \\
\hline\hline
\multirow{2}{*}{\; $r$ (kpc) \cite{Rubin1980} \;} & \multirow{2}{*}{\; ${v_\varphi}_{msr}$ (km/s) \cite{Rubin1980} \;} & \multirow{2}{*}{\; ${v_\varphi}$ (km/s) \;} & \multirow{2}{*}{\; $s_e$ (km/s) \;} & \multicolumn{3}{c}{Confidence interval} \\ \cline{5-7}
& & & & $68\%$ & $95\%$ & $99\%$ \\ \hline\hline
1.5 & 123 & 146 & 14 & $ \; [132, 161] \; $ & $ \; [117, 176] \; $ & $ \; [106, 186] \; $ \\ \hline
2 & 154 & 165 & 14 & $ \; [151, 180] \; $ & $ \; [136, 195] \; $ & $ \; [126, 205] \; $ \\ \hline
2.5 & 177 & 177 & 14 & $ \; [162, 191] \; $ & $ \; [147, 206] \; $ & $ \; [137, 217] \; $ \\ \hline
3 & 193 & 185 & 14 & $ \; [170, 199] \; $ & $ \; [155, 214] \; $ & $ \; [145, 225] \; $ \\ \hline
4 & 218 & 194 & 14 & $ \; [180, 209] \; $ & $ \; [165, 224] \; $ & $ \; [154, 234] \; $ \\ \hline
5 & 232 & 200 & 14 & $ \; [185, 215] \; $ & $ \; [170, 230] \; $ & $ \; [160, 240] \; $ \\ \hline
6 & 228 & 204 & 14 & $ \; [189, 218] \; $ & $ \; [174, 233] \; $ & $ \; [164, 244] \; $ \\ \hline
7 & 216 & 207 & 14 & $ \; [192, 221] \; $ & $ \; [177, 236] \; $ & $ \; [166, 247] \; $ \\ \hline
8 & 208 & 209 & 14 & $ \; [194, 223] \; $ & $ \; [179, 238] \; $ & $ \; [168, 249] \; $ \\ \hline
9 & 212 & 210 & 14 & $ \; [196, 225] \; $ & $ \; [181, 240] \; $ & $ \; [170, 250] \; $ \\ \hline
10 & 220 & 211 & 14 & $ \; [197, 226] \; $ & $ \; [182, 241] \; $ & $ \; [171, 252] \; $ \\ \hline
12 & 230 & 213 & 14 & $ \; [199, 228] \; $ & $ \; [184, 243] \; $ & $ \; [173, 254] \; $ \\ \hline
14 & 213 & 215 & 14 & $ \; [200, 229] \; $ & $ \; [185, 244] \; $ & $ \; [175, 255] \; $ \\ \hline
16 & 218 & 216 & 14 & $ \; [201, 230] \; $ & $ \; [186, 245] \; $ & $ \; [176, 256] \; $ \\ \hline
18 & 225 & 217 & 14 & $ \; [202, 231] \; $ & $ \; [187, 246] \; $ & $ \; [176, 257] \; $ \\ \hline
20 & 228 & 217 & 14 & $ \; [203, 232] \; $ & $ \; [188, 247] \; $ & $ \; [177, 257] \; $ \\ \hline
22 & 226 & 218 & 14 & $ \; [203, 232] \; $ & $ \; [188, 247] \; $ & $ \; [178, 258] \; $ \\ \hline
24 & 220 & 218 & 14 & $ \; [204, 233] \; $ & $ \; [188, 248] \; $ & $ \; [178, 258] \; $ \\ \hline
26 & 209 & 219 & 14 & $ \; [204, 233] \; $ & $ \; [189, 248] \; $ & $ \; [178, 259] \; $ \\ \hline
28 & 203 & 219 & 14 & $ \; [204, 234] \; $ & $ \; [189, 249] \; $ & $ \; [179, 259] \; $ \\ \hline
30 & 204 & 219 & 14 & $ \; [204, 234] \; $ & $ \; [189, 249] \; $ & $ \; [179, 259] \; $ \\ \hline
32 & 212 & 219 & 14 & $ \; [205, 234] \; $ & $ \; [190, 249] \; $ & $ \; [179, 260] \; $ \\ \hline
34 & 208 & 220 & 14 & $ \; [205, 234] \; $ & $ \; [190, 249] \; $ & $ \; [179, 260] \; $ \\ \hline
38 & 202 & 220 & 14 & $ \; [205, 235] \; $ & $ \; [190, 250] \; $ & $ \; [180, 260] \; $ \\ \hline
42 & 204 & 220 & 14 & $ \; [206, 235] \; $ & $ \; [191, 250] \; $ & $ \; [180, 260] \; $ \\ \hline
46 & 208 & 220 & 14 & $ \; [206, 235] \; $ & $ \; [191, 250] \; $ & $ \; [180, 261] \; $ \\ \hline
47.4 & 211 & 221 & 14 & $ \; [206, 235] \; $ & $ \; [191, 250] \; $ & $ \; [180, 261] \; $
\end{tabular}
\label{NGC_801_CI}
\end{center}
\end{table}


\begin{thebibliography}{99}

\bibitem{Brans1961} C. Brans and R. H. Dicke, Phys. Rev. \textbf{124}, 925 (1961).

\bibitem{limite_experimental} B. Bertotti, L. Iess and P. Tortora, Nature \textbf{425}, 374 (2003); M. W. Clifford, Living Rev. Relativity \textbf{9}, 3 (2006).

\bibitem{Damour1993} Th. Damour and K. Nordtvedt, Phys. Rev. Lett. \textbf{70}, 2217 (1993).

\bibitem{Damour1993A} Th. Damour and K. Nordtvedt, Phys. Rev. D \textbf{48}, 3436 (1993).

\bibitem{Polyakov1994} Th. Damour and A. M. Polyakov, Nucl. Phys. B \textbf{423}, 532 (1994).

\bibitem{Contaldi1999} C. Contaldi, M.
B. Hindmarsh and J. Magueijo, Phys. Rev. Lett. \textbf{82}, 2034 (1999).

\bibitem{Zwicky1933} F. Zwicky, Helvetica Physica Acta \textbf{6}, 110 (1933).
%die rotverschiebung von extragalaktischen nebeln
%110-127

\bibitem{Zwicky1937} F. Zwicky, The Astrophysical Journal \textbf{86}, 217 (1937).
%on the masses of nebulae and of clusters of nebulae

\bibitem{Rubin1980} V. C. Rubin, W. K. Ford Jr. and N. Thonnard, The Astrophysical Journal \textbf{238}, 471 (1980).

\bibitem{Persic1995} M. Persic and P. Salucci, Ap. J. Suppl. Ser. \textbf{99}, 501 (1995).

\bibitem{Persic1997} P. Salucci and M. Persic, Proc. of the Sesto DM1996 Conference (1997).

\bibitem{Guzman2002} T. Matos, F. S. Guzmán, L. A. Ureña-López and D. Núñez, \textit{Scalar field dark matter}, in Exact Solutions and Scalar Fields in Gravity -- Recent Developments, Edited by A. Macias, J. L. Cervantes-Cota and C. Lämmerzahl, 165, Kluwer Academic Publishers (2002).

\bibitem{Lee2004} Tae Hoon Lee e Byung Joo Lee, Phys. Rev. D \textbf{69}, 127502 (2004).

\bibitem{Leineker2006} M. Leineker Costa, A. L. Naves de Oliveira and M. E. X. Guimar\~aes, Int. J. Mod. Phys. D \textbf{15}, 387 (2006).

\bibitem{Jensen1992} B. Jensen, Class. Quantum Grav. \textbf{9}, L7 (1992).

\bibitem{artigo_corda_reta} S. Mittmann Santos, J. M. Hoff Silva and J. L. Cindra, \textit{Straight spinning cosmic strings in Brans-Dicke gravity}, Submitted for publication, arXiv: 1708.00507.

\bibitem{Deser1992} S. Deser and R. Jackiw, Comments Nucl. Part. Phys. \textbf{20}, 337 (1992).

\bibitem{Mazur1986} P. O. Mazur,
%\textit{Spinning cosmic strings and quantization of energy},
Phys. Rev. Lett. \textbf{57}, 929 (1986).

\bibitem{Kostelecky1989} V. A. Kostelecký and S. Samuel, Phys. Rev. D \textbf{39}, 683 (1989).

\bibitem{Kostelecky1991} V. A. Kostelecký and R. Potting, Nucl. Phys. B \textbf{359}, 545 (1991); V. A. Kostelecký and R. Potting, Phys. Rev. D \textbf{51}, 3923 (1995); V. A. Kostelecký and R. Potting, Phys. Lett. B \textbf{381}, 89 (1996).

\bibitem{Faraoni1999} V. Faraoni, E. Gunzig and P. Nardone, Fundam. Cosmic Phys. \textbf{20}, 121 (1999).

\bibitem{Quiros2013} I. Quiros, R. Garcia-Salcedo, J. E. M. Aguilar and T. Matos, Gen. Rel. Grav. \textbf{45}, 489 (2013).

\bibitem{Boisseau1998} B. Boisseau and B. Linet, Gen. Rel. Grav. \textbf{30}, 963 (1998).

%\bibitem{Soleng1990} H. H. Soleng, Class. Quantum Grav. \textbf{7}, 999 (1990).

%\bibitem{Plebanski2006} J. Pleba\'nski and A. Krasi\'nski, \textit{An Introduction to General Relativity and Cosmology}, %Cambridge University Press (2006).

\bibitem{Godel1949} K. G\"odel, Rev. Mod. Phys. \textbf{21}, 447 (1949).

\bibitem{Matos2014} T. Matos, D Nuñez, F. S. Guzmán e E. Ramírez, Gen. Rel. Grav. \textbf{34}, 283 (2002).

\bibitem{Chandrasekhar1983} S. Chandrasekhar, \textit{The Mathematical Theory of Black Holes}, Oxford University Press (1983).

\bibitem{Hawking1992} S. W. Hawking, Phys. Rev. D \textbf{46}, 603 (1992).

\bibitem{Thorne1988} M. S. Morris, K. S. Thorne and U. Yurtsever, Phys. Rev. Lett. \textbf{61}, 148 (1988); J. Friedmann, M. S. Morris, I. D. Novikov, F. Echeverria, G. Klinkhammer, K. S. Thorne and U. Yurtsever, Phys. Rev. D \textbf{42}, 1915 (1990).

\bibitem{Novello1992} M. Novello, N. F. Svaiter and M. E. X. Guimar\~aes, Mod. Phys. Lett. {\bf A7}, 381 (1992).

\bibitem{Novello1993} M. Novello, N. F. Svaiter and M. E. X. Guimar\~aes, Gen. Rel. Grav. {\bf 25}, 137 (1993).

\bibitem{Novello1994} M. Novello and M. C. M. da Silva, Phys. Rev. D \textbf{49}, 825 (1994).

\bibitem{gravitacao_quantica} S. W. Hawking, \textit{The path integral approach to quantum gravity}, in General
Relativity, an Einstein Centenary Survey, Edited by S. W. Hawking and W. Israel, Cambridge University Press (1979); J. B. Hartle and S. W. Hawking, Phys. Rev. D \textbf{28}, 2960 (1983).

%\bibitem{Gregory1988} Ruth Gregory, Phys. Lett. B \textbf{215}, 663 (1988).

%\bibitem{Spiess2010} A.-N. Spiess and N. Neumeyer, \textit{An evaluation of $R^2$ as an inadequate measure for nonlinear %models in pharmacological and biochemical research: a Monte Carlo approach}, BMC Pharmacology \textbf{10}:6, number 1 %(2010).

\end{thebibliography}
\end{document}